# Advances in Atomic Time Scale imaging with a Fine Intrinsic Spatial Resolution

Jingzhen Li[1]*, Yi Cai[1], Xuanke Zeng[1], Xiaowei Lu[1], Qifan Zhu[1], and Yongle Zhu[1]

*Shenzhen Key Laboratory of Micro/Nano Photonic Information Technology, Institute of Photonic Engineering, School of Physics and Opto-electronic Engineering, Shenzhen University, Shenzhen 518060, China*

*Corresponding author. Email: lijz@szu.edu.cn

**Abstract**

Atomic time scale imaging, opening a new era for studying dynamics in microcosmos, is presently attracting immense research interesting on the global level due to its powerful ability. On the atom level, physics, chemistry, and biology are identical for researching atom motion and atomic state change. The light possesses twoness, the information carrier and the research resource. The most fundamental principle of this imaging is that light records the event modulated light field by itself, so called all optical imaging. This paper can answer what is the essential standard to develop and evaluate atomic time scale imaging, what is the optimal imaging system, and what are the typical techniques to implement this imaging, up to now. At present, the best record in the experiment, made by multistage optical parametric amplification (MOPA), is realizing 50 fs resolved optical imaging with a spatial resolution of ~83 lp/mm at an effective framing rate of $15 \times 10^{12}$ fps for recording an ultrafast optical lattice with its rotating speed up to $13.5 \times 10^{12}$ rad/s.

## 1. Introduction

Dr. F. Krausz [1,2] expounded on the relationship between energy, motion, and time category in microcosmos, that is, the different microscopic motion corresponds to the definite time scale and certain energy region. In general, physics, chemistry, and biology are identical on the atom level, and all of them research atom motion and atomic state change, as shown in table 1, the characteristic time of which is about picosecond to femtosecond, so called atomic time scale.

It has always been the dream of scientists to observe the light information of the atomic time process, accurately reveal the physical, chemical, and biological view and its evolution rules in the process of atomic movement, and promote the progress of human science. Atomic time scale imaging with a fine intrinsic spatial resolution is applicable for studying the ultrafast transient events generated in ultrafast physics, ultrafast chemistry, and ultrafast biology, such as ultrafast photophysical processes of semiconductor and quantum well microstructures, photoexcited excitons, and carriers excitation, high order harmonic effects [3~6], and super-high-strength laser wake-field acceleration and ultrafast dynamics processes of condensed state materials [7~10];

formation and fracture of chemical bonds, transfer of protons and electrons, and effects of vibration and rotation in molecules on chemical reactions; energy transfer process in photosynthesis, the photoisomerization process in the visual system, and the charge transfer and proton transfer process in DNA, and so on.

The most fundamental principle of this imaging is recording the event modulated light field by light itself, so called all optical imaging. It is based on the light twoness: the information carrier and the research resource which includes light velocity and photon itself, and its amplitude, phase, wavelengths, and polarization, and the high order nonlinear optical effects and the quantum property of photon. The atomic time scale imaging has been developed through half a century; only in this century, its rapid development has been achieved.

Table 1. Atomic time scale

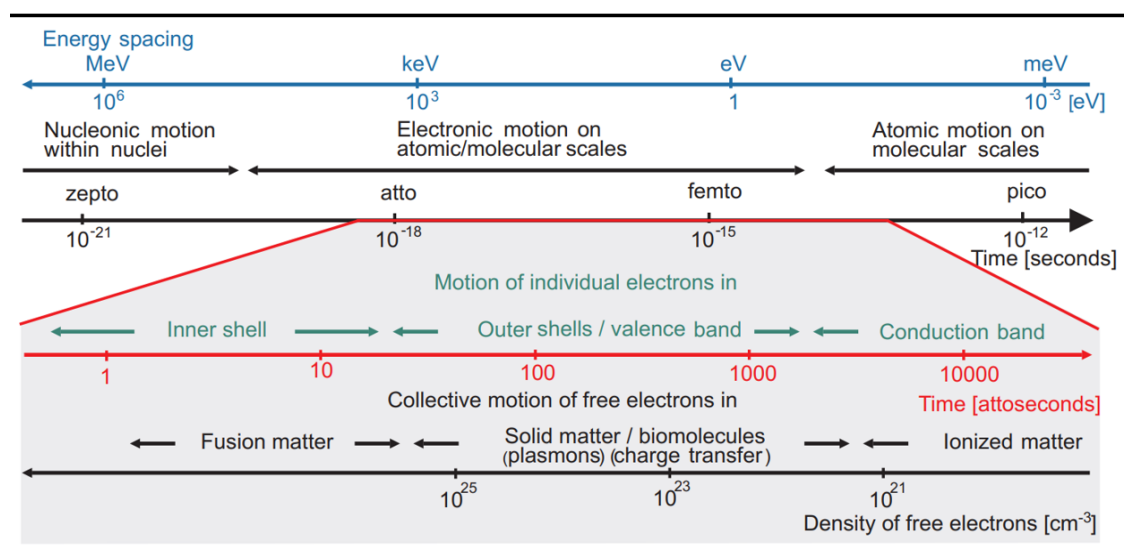

Fifty years ago, Dr. M. A. Duguay [11,12], a pioneer in the atomic time scale imaging field, published the paper named "Ultrahigh speed photography of picosecond light pulses and echoes", which may be the first paper about picosecond imaging by means of the light-induced Kerr effect of $CS_2$ liquid. A giant laser pulse-driven Kerr cell camera can freeze the green pulse propagation, and adopting the steep front edge of the laser pulse can further shorten the shutter time to some 0.1-1 ps for the laser pulse width of 8 ps. The limit of the shutter time depends on the relaxation time of the Kerr medium. For the better Kerr medium such as $CCl_4$ and optical glass BK7, the relaxation time related to the nonlinear Kerr index $n_{2B}$ is about 0.5 ps, less than 10 fs, respectively. After that, it has been to continue enhancing the three-order nonlinear effect, seeking shorter relaxation time Kerr medium, and making the light Kerr shutter and its array to record more pictures in a single shot [13-17].

About forty years ago, another atomic time scale imaging pioneer, Dr. Nils Abramson [18-20], first presented the holographic coherent shutter conception and made a light-in-flight recording of a holographic motion picture of ultrafast phenomena. The connotation of this shutter is that if the optical path difference between the object light beam and the reference light beam is less than the coherent length, the coherent shutter is open to record coherently; otherwise, the shutter is closed. Here, the coherent length depending on the ultra-short laser pulse width, is used to record the light-wave propagation process.

The light-in-flight recording has the ellipsoidal cluster for every point on the sensitive plate responding to a zero optical path difference and the hyperboloid cluster for every point on the object plate. If the coherent length is very shorter, the partial object intersected by the bright fringe of zero optical path difference can only be observed when reconstructing. There were many experiments made by light-in-flight recording at the framing rate of $10^{10}$ fps, such as the wavefront of light reflected by a mirror and focused by a lens and light passing through interferometers. After that, the framing rate of $10^{12}$ fps, using the just developed colliding mode-locked dye laser, was reached in China [22], the propagation process of a picosecond light pulse diffracted in diffuse transmission medium was recorded by Japanese scholars [23], and the propagation process of a femtosecond light pulse was recorded [24-27].

This imaging technique, unifying the shuttering and framing functions, is more available to study ultra-short laser pulse transmission and luminescence very cleanly and sharply. A femtosecond laser pulse moving along the film plane plays a role in coherent pulse-gating. With the femtosecond laser pulse technique developing further, it would be possible to unify the higher temporal and fine spatial resolution to research microcosmos with an exposure time of $10^{-15}$ s or less and a framing rate of $10^{14}$ fps or higher.

In 1999, American scientist A. H. Zewail won the Nobel Prize for his studies of the transition states of chemical reactions through multi-plicating pump-probe. He is the first time observing the excited state, the covalent bond-breaking and ionic state-forming of a chemical reaction by the experiment of the bond-breaking process of ICN to open a new era of femtosecond imaging [28]. This multi-plicating pump-probe method, whose original schematic diagram is shown in Figure 1, is suitable for studying the transient phenomena of periodic repetition.

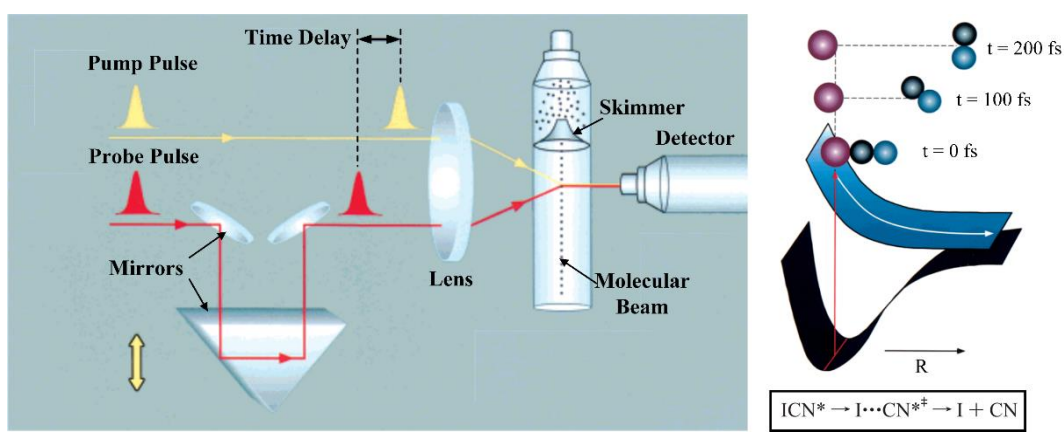

**Figure 1. The original schematic diagram of a bond-breaking process of ICN.** Experiment set and result explaining; multi-plicating pump-prob: time-delay 10 fs between adjacent two pumps

Since the turn of the century, the development of ultrafast laser technology has pushed the time resolution of pump-detection technology into the femtosecond - attosecond region, and many events in physics, chemistry, and biology, have been studied by means of this technique [29-35]. But, the pump-probe cannot study difficult-to-reproduce, non-repetitive events, probabilistic, or complex events, such as explosions, destructive testing, quantum-mechanical processes, Brownian motion, shockwave interaction in living cells, protein folding, enzyme reactions and thermal dynamics in semiconductors. These events can only be done by single-shot ultrafast imaging methods, which have made great progress and emerge continuously new imaging principles and technologies.

Laser-induced optical waveguide deflection scanning imaging is an all-optical solid-state scanning camera with a time resolution of 2.5 ps and a dynamic range of more than 3000 [36]. K. L. Baker proposed all-optical solid framing imaging [37]. Z. Li proposed single frequency domain tomography (FDT), which successfully achieved 5 pictures at a framing rate of $4.2 \times 10^{11}$ fps, but with a frame format of 16×10 pixels [38]. L. Gao proposed compression sensing ultra-fast imaging technology (CUP), which can achieve ~~5 pictures~~ 350 frames at a framing rate of $10^{11}$ fps with a spatial resolution of 0.78 lp /mm through a digital micromirror array for spatial coding, compressed-sensing-based algorithm and convert tube streak camera[39,40]. CUP may create a new area of computed femtosecond imaging. K. Nakagawa proposed time-series all-optical imaging technology (STAMP), which once attracted academic attention and achieved 6 images of a frame format of 450 ×450 pixels at a framing rate of $4.4 \times 10^{12}$ fps correspondent to the framing time of 299 fs, and the exposure time was 733 fs [41].

Currently, some new imaging principles and technologies have emerged. One of the novel technologies is multistage optical parametric amplification (MOPA), which can realize 50 fs -resolved optical imaging at an effective framing rate of $15 \times 10^{12}$ fps to record an ultrafast optical lattice with its rotating speed up to $13.5 \times 10^{12}$ rad/s [42]. The frequency domain integration sequential imaging (FISI), encoding a transient event by an inversed 4f system and decoding it using optical frequency recognition, achieved shadow imaging of the plasma channel with an intrinsic spatial resolution of 108 lp/mm at a framing rate of $12.6 \times 10^{12}$ fps [43]. A novel imaging with the raster principle (OPR) that can capture real-time imaging of ultrafast dynamics is demonstrated with the spatial resolution of ~90 lp/mm, the frame number of 12 at a framing rate of 2 trillion fps [44]. Up to now, there are at least six kinds of representative atomic time scale imaging to be emphatically expounded in principle, property, and characteristics. They are femtosecond holography, compressed ultrafast photography with the convert tube, femtosecond imaging coded on spectrum plan, spectrally coded femtosecond imaging, raster-coded femtosecond imaging, and idler light femtosecond imaging.

## 2. High-speed imaging information theory

High-speed imaging information theory can be used to evaluate high-speed imaging systems and techniques and how to thoroughly upgrade the performances and exploit the resources of the recording light used as an information carrier.

C. E. Shannon, a great pioneer, created information theory: communication engineering having nothing to do with semantics, information being random in nature, and formalization and randomization. Shannon's main contribution to optical information theory is accurately proposing the quantitative description of information, information entropy, and sampling theorem. G. Toraldo di Francia, another pioneer in optical information theory, introduced the concept of optical degree of freedom and ellipsoidal wave function into optics and established the theory of optical eigenvalue, and deduced a perfect set of integral equations, including the complete orthogonality equation, the integral imaging equation and the characteristic equation of Fourier transform [45-47]. M. V. Lauve and W. Lukosz founded that Laue-Lukosz invariant principle of degrees of freedom, which is the fundamental principle that the total amount of information in an optical system is conservative [48,49].

*2.1. Schardin temporal-spatial information equation.*

H. Schardin [50], a German scientist, a pioneer in high-speed photography, based on the optical information theory, deduced Schardin temporal-spatial information equation and established the high-speed photography information theory covering temporal information, spatial information, temporal modification factor $g$, spatial information aliasing ratio $a_s$, intrinsic spatial resolution $n$, and temporal resolution $\Delta t_{opt}$, to guide researching and evaluating the performance of high-speed photography.

This equation can be expressed as follows

$$\Im = \Im_R \Im_T = F \cdot n^2 \cdot \ln \kappa \cdot f_\omega \cdot g^{2/3}$$
$$\Im_T = f_\omega \cdot g^{2/3} \quad (1)$$
$$\Im_R = F \cdot n^2 \cdot \ln \kappa$$

Where $\Im$ is the total amount of spatio-temporal information of a high-speed photography system called the spatio-temporal information rate, that is, the number of bits transmitted and recorded by a high-speed camera per second, and is the most basic basis for evaluating the performance of all kinds of ultra-high speed photography technology and has universal applicability. This rate consists of time information $\Im_T$ and space information $\Im_R$.

In the expression of time information $\Im_T$, the framing rate $f_\omega$ of the high-speed camera is the basic parameter to characterize the time resolution of the camera. The framing rate, that is, the imaging rate, is the reciprocal of the time interval between two adjacent images. To accurately describe the amount of time information, H. Schardin derived the time information correction factor $g^{2/3}$. $g$ is the time information quality factor and equal to the ratio of framing time $t_f$ to effective exposure time $t_e$. The spatial information amount $\Im_R$ is the product of the spatial bandwidth product $F \cdot n^2$ of a single picture and the information amount $\ln \kappa$ of a single pixel (that is single channel), where $\kappa = 1+m$ is generally signal-to-noise ratio, $m$ is information level, $F = b \times h$ is the area of the picture, $b$ is the width of the time direction of it, $h$ is the height of the spatial direction of it, $n^2$ is the spatial resolution of it (when the spatial resolution of the time and spatial directions is equal).

*2.2. Spatial information aliasing ratio $a_s$, intrinsic spatial resolution n, and temporal resolution $\Delta t_{op}$.*

These parameters are the most important and fundamental in the high-speed imaging field, especially in the femtosecond imaging. Spatial information aliasing ratio $a_s$ is defined as the ratio of the overlapped area of exposure $A_{overlap}$ to exposure area $A_0$, which has a close relation with time information quality factor $g$ and can accurately state the extent to which the image quality is affected.

As a matter of fact, if the time information quality factor $g$ is far less than 1, the information confusion, described by the spatial information aliasing ratio $a_s$, is extremely serious, and its analysis value is not much.

The intrinsic optical spatial resolution $n$ of the imaging system on the image plane is the function of the spatial resolution of diffraction, that of the high-speed formation principle, and that of the imaging sensor. And their reciprocals are the inherent spatial blur amount $\sigma$, and the spatial blur $\sigma_d$, $\sigma_h$, and $\sigma_s$, respectively; and the relationship between them can be expressed by

$$\sigma = f(\sigma_d, \sigma_h, \sigma_s, \cdots) \quad (2)$$

Generally, the inherent space blur amount $\sigma$ can be synthesized using the error accumulation law. The intrinsic optical spatial resolution $n$ can be written as $n = 1/\sigma$. However, the optical spatial resolution of the imaging system on the image plane and the pixel number/pixel size of charge coupled device (CCD) are two completely different concepts, and the optical spatial resolution is dependent on multi-factors apart from the size of a pixel of a photosensitive element of CCD. Russian scholars had done in-depth test research on the scientific grade CCD(ST-71): the half-peak width, that is, the inherent spatial blur amount $\sigma_s$ of the array CCD itself, is 1.5 times of the distance between adjacent CCD pixels in the visible light region and 2 times of the distance in the infrared band [51].

The definition of temporal resolution $\Delta t_{opt}$ is an ability to separate two adjacent temporal states of transient event: the effective exposure time for single frame high-speed imaging, the framing time modified by time information factor for high-speed framing imaging, and the discernible shortest resolution time for high-speed scanning imaging which is time equivalent of convolution of slit function rect $(-x/b)$ and system line spread function $L(x)$ and can be equal to the system's slit image blur divided by the maximum scanning speed $V_{max}$ of the system, in magnitude. The corresponding slit image function and temporal resolution equation are as follows

$$E(x) = \text{rect}(-\frac{x}{b}) \otimes L(x)$$
$$\Delta t_{opt} = \frac{\sigma(E(x))}{V_{max}} \quad (3)$$

where $E(x)$ is a slit image function.

*2.3 three fundamental principles of atomic time scale imaging*

Our group has been studying ultra-high speed imaging, especially atomic time scale imaging, for a long time and always exploring what is the basic rule and what is the best optimum imaging system in this circle. The preliminary conclusion is that:

The first principle: Schardin temporal-spatial information rate is the most basic, reliable, and essential standard to develop and evaluate high sped imaging, deduced from Laue-Lukosz invariant principle under the high-speed situation, which is universal in the optical information field.

The second principle: pursuing the optimal high-speed photography systems is more important, the connotation of which is that framing time, exposure time, spatial resolution, and frame size are irrelated to each other, and no limit by Heisenberg uncertainty principle.

The third principle: the shorter exposure time is the most critical factor for increasing the information rate, which is dependent on the ultrafast laser pulse width, further compression of which is always on the road.

## 3. Femtosecond holography

Since Light-in flight recording, first presented by Nils Abramson, is one of holography with the holographic coherent shutter, the atomic time scale holography has been rapidly developed, and great achievements have been made by scientific workers around the world [52-54]. Holography was originally used for single-shot imaging of the femtosecond process benefiting from its multiple parameters and imaging coding to be easy. The scalar wave equation is that

$$R = R(x,y,x)\exp\{-i\frac{2\pi}{\lambda}[\xi x+\eta y+\zeta z]+i\varphi(x,y,z)\} \qquad (4)$$

where $R(x, y, z)$, $\lambda$, $\xi$, $\eta$, $\zeta$, and $\varphi(x, y, z)$ can be used as encoding parameters. Moreover, more and more encoding parameters can be used from the vector wave equation, such as polarization, spin, and orbital angular momentums applied to OAM multiplexing nonlinear holography [55] and polarization degree of the entangled state applied to quantum holography [56].

*3.1. Time domain holography (TDH)*

Using quasi-coaxial holography combined with amplitude splitting and coding of bearing angle realized the four-frame holography of the nonlinear process with femtosecond time resolution, as shown in Figure 4, may be the first of time domain holography in the atomic time scale [57,58].

In Figure 2 the four mirror segments are the key units for the wavefront splitting, the framing, and the high-speed forming, each of which has independent controls for an angle to adjust the propagation direction of the sub-pulses for the bearing angle coding and for axial displacements to adjust the relative time delay between them for the time sequential recording. As a result, the four probing pulses have been made to spatially overlap in the interaction region and then spatially separate on the recording plane. The hologram with the temporal resolution of 150 fs and the spatial resolution of 4 μm have been recorded.

Since then, the femtosecond time domain holography has made good progress; in which the typical ones are the ultra-fast digital holography of femtosecond order with solid angle encoding made [59,60] and all-optical coaxial framing holography with the exposure time of 10ps and the framing time of 34 ps using parallel coherence shutter made by Dr. Chen Guanghua [61].

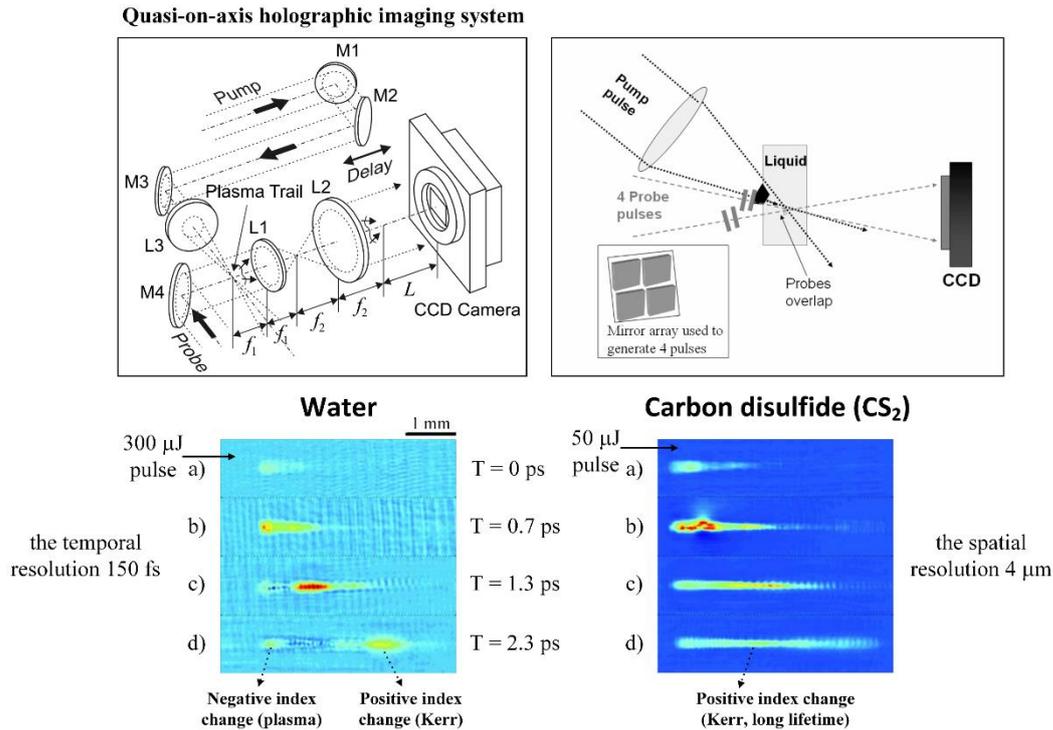

**Figure 2. Original femtosecond holography made by Dr. Martin Centurion.** Imaging system diagram (up), holograms of water and CS2 (down)

*3.2. Frequency domain holography (FDH)*

Frequency-domain holography combines chirped pulse technology with frequency-domain interference technology to obtain transient disturbance information during a period of the ultrafast change process through only one interferometry.

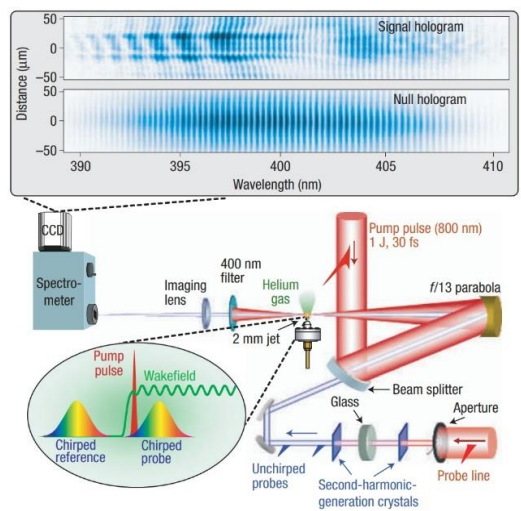

**Figure 3: Plasma wake-field recorded by FDH.** Its system diagram(down), a frequency-domain hologram of plasma wakefield (up)

In Figure 3, there is the first frequency-domain hologram to record laser-induced plasma wake-field using a chirped-pulse and spectral interferometry, in which each wavelength corresponds to a certain moment for transforming the temporal profile of an incident light into a spectral profile on an image sensor to primarily observe plasma wake-field wavefront curvature with 30 fs temporal resolution [62,63].

Single-shot tomographic movies of evolving light-velocity object is another novel achievement in frequency-domain holography with a frame number of 5 frames, the framing rate of $4.2 \times 10^{11}$ fps, and the spatial resolving power of 20 μm but the mean squared error of 0.5 [38].

a) *Sparsely sampled frequency domain holography (SSFDH)*

This is the computing compressed frequency domain holography with sparsely sampled data based on the sparse original signal in the frequency domain. The image frequency signal is more concentrated near the central fundamental frequency, the pseudo-random sampling template is optimizing the encoding method and the filling rate, and the minimum constraint model is constructed. The iterative method is optimized to recover the optimal initial distribution.

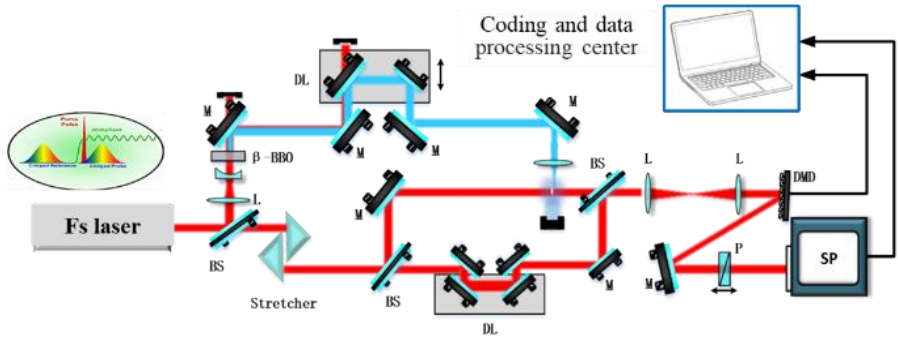

**Figure 4: Schematic diagram of the experimental setup.** DMD pseudo-random sampling template, SP spectral interferometry, Stretcher producing a chirped-pulse

SSFDH schematic diagram of the experimental setup is in Figure 4. There are three key elements: 1) the spectral interferometry SP for transforming the temporal profile of an incident light into a spectral profile on an image sensor and realizing interfere between objective frequency and reference frequency; 2) the pseudo-random sampling template SP for the spatial encoding to realize two dimension recording instead of one dimension recording; 3) the stretcher for producing a chirped-pulse to realize each wavelength corresponding to a certain moment of the event.

Sparsely sampling compression recording and reproduction can be explained in this way. Sparsely sampling compression recording equation is

$$H = \sum_Y I(u+u_Y, v) \cdot b(u+u_Y, v) \qquad (5)$$

where $b(u+u_Y, v)$ is a pseudo-random sampling template function, $I(u+u_Y, v)$ is each individual original image function, $H$ expresses stacking sample images from which each individual image cannot be directly recovered. In order to recover images from the experimental result, the minimum constraint function should be constructed, and reproducing calculation should be performed

$$\min_{x}[f(x_1, x_2, \ldots, x_n)] = \min_{x}\left[\frac{1}{2}\|y - (B_1x_1 + B_2x_2 + \cdots + B_nx_n)\|^2 + \lambda(\Theta(x_1) + \Theta(x_2) + \cdots + \Theta(x_n))\right] \quad (6)$$

where $y$ represents recorded information, $\lambda$ is an adjustment parameter, $x$ is original information, $B$ is equivalent to $b$, $\Theta(x)$ is an original signal function of sparse constraint with DMD sampling template which is here designed at the fine filling rate of 0.08. As a result, in our experiment, the optimal initial distribution of the evolution process of laser-induced plasma can be recovered through the iterative method to optimize. According to the pre-calibration coefficient, the carrier phase was solved by FFT, the spatial phase distribution of XYT was obtained by subtraction calculation, and the minimum time resolution of Fourier transform method, limited by the spectral bandwidth of the reference pulse, is 71.13fs at a single pixel corresponding time scale of 24fs. Sparsely sampled frequency domain holography in atomic time scale is the fine way to study the dynamics process of atom and molecule, but there is a long way to go for its accuracy and optimal scope of application.

## 4. Compressed ultrafast photography (CUPs)

*4.1. CUPs imaging including CUP, T-CUP, CUSP and several derivative articles*

The compressed ultrafast photography (CUP), presented by Dr. Liang Gao in the paper [39], can capture non-repetitive time-evolving events as ultrafast luminescent phenomena with a two-dimensional frame, which produced a sensational effect in the imaging world and opened a new way to realize femtosecond imaging with a compressed sensing algorithm.

CUP system configuration and schematic diagram of tube streak camera are in Figure 5. In the light of compressed sensing, CUP can be carried out through encoding the spatial domain with a pseudo-random template, shearing in the temporal domain by a streak camera with a fully opened entrance slit, imaging the encoded and sheared three-dimensional (3D) $x$, $y$, $t$ scene on a CCD, and reconstructing images by iterative estimation of a solution of minimizing an objective function. There are three key techniques: digital micromirror array DMD for spatial coding, compressed sensing based algorithm, and convert tube streak camera with Hamamatsu C7700 (the temporal resolution of 5 ps and the maximum streak velocity of 26.4 mm/ns, binned pixel of $2d = 13$ μm) for widening slit shearing. The main specifications are the maximum framing rate of $10^{11}$ fps, the spatial resolution of 0.36～0.43 lp/mm (0.78 lp/mm in unshearing), and the exposure time of 10 ps.

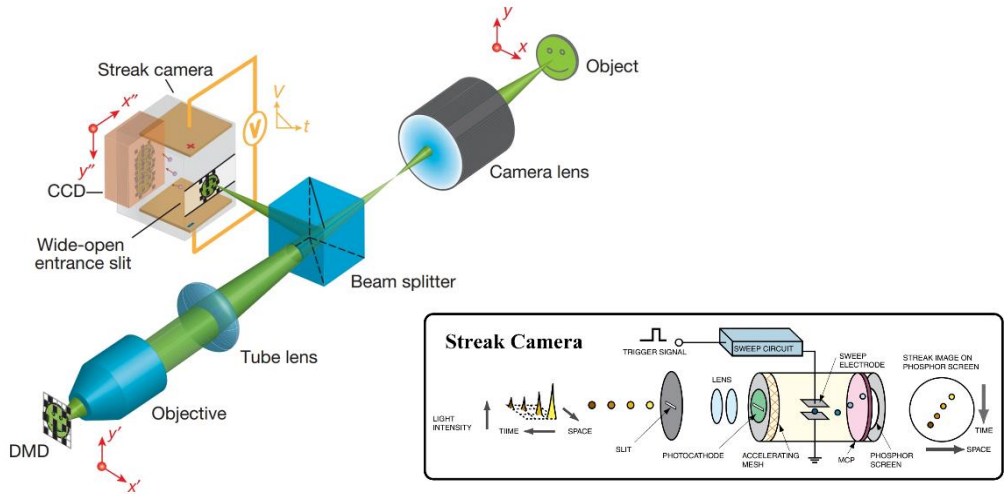

**Figure 5. CUP system.** Schematic diagram of CUP system and tube streak camera

The feature of T-CUP [64], formed based on CUP, is a double view regression data processing method, an added external CCD camera, and the tube streak camera with a C6138 with the temporal resolution of 200 fs, which is not essentially different from CUP. The maximum framing rate of $10^{13}$ fps was calculated by a one-pixel size of 6.45 μm divided by the maximum streak velocity of 64.5 mm/ns.

As to CUSP [65], its characteristic is an added spectrum coding with grating for increasing framing rate and the pulse subdivision for increasing frame number based on T-CUP. The maximum framing rate is $70×10^{12}$ fps in an active model and $0.5×10^{12}$ fps in a passive model. And this calculation is based on the scene sheared by one pixel per frame in the vertical direction, so the time interval between adjacent frames is V/d.

*4.2 Preliminary studying and evaluating CUPs imaging*

CUPs, including CUP, T-CUP, CUSP, and several derivative systems [66,67], is a kind of computational imaging technology based on prior information through pseudo-random template for coding, tube streak camera for shearing and compression sensing, and are mainly used for the study of ultra-fast phenomena of laser pulse transmission and luminescence. Since 2014, our group has been studying the image quality, spatial resolution, framing rate, and frame number of CUPs imaging, and the preliminary results are as follows.

1) Optical spatial resolution (not the pixel size of CCD) is not well for study the details. Test data in CUP would be reliable, 0.36 ~ 0.43 lp/mm (Unshearing 0.78 lp /mm); T-CUP did not give the measured values, only the comparison patterns under several measured conditions; CUST did not involve spatial resolution. The reason for the poor spatial resolution is from the image tube streak camera: principle error of the image tube cameras: three times conversion of light changed to electricity, and to light, at last to charge; two-dimensional of scanning slit to increase aberration in the electronic-optical system; shearing to cause a dynamic error, further reducing the spatial resolution.
2) The calculated effective framing rate is worth further discussion. This is a question of how to calculate the framing rate.

It is not appropriate to divide the binned pixel size (2× or 1× CCD pixel size in 3 articles mentioned above) by the scanning speed of the streak camera, but would rather appropriate to divide the imaging blur σ of the tube streak camera, which is much larger than the CCD pixel size and equal to reciprocal of the spatial resolution of the whole system.

The effective framing rate should be based on the performance characteristics of the used streak cameras in CUPs. For example, the streak camera used by CUP is Hamamatsu C7700 with the maximum scanning velocity of 26.4 mm ns$^{-1}$, 5 ps temporal resolution, and 6.5 μm pixel size. In the original text, the CUP operates at $10^{11}$ fps with the streak camera's scanning velocity set to 1.32mm ns$^{-1}$. This means that the imaging blur is 13.2 μm (1.32 mm ns$^{-1}$ / (1/$10^{11}$fps)), accounting for only 2× pixel size. However, the imaging blur of Hamamatsu C7700 should be 0.132 mm (26.4 mm ns$^{-1}$×5ps), accounting for about 20× pixel size. So one can get the effective framing rate should be $10^{10}$ fps (1/(0.132mm/1.32 mm ns$^{-1}$)) for getting clear image instead of fuzzy image. The same method can be used to calculate the effective framing rate in T-CUP and CUSP. After calculating, the corrected specifications are as follows: CUP's effective framing rate should be $10^{10}$ fps, not $10^{11}$ fps; T-CUP's effective framing rate should be 0.65×$10^{12}$ fps, not 10×$10^{12}$ fps; CUSP's effective framing rate should respectively be 4.52×$10^{12}$ fps for active, not 70×$10^{12}$ fps; 3.23×$10^{10}$ fps for passive, not 0.5×$10^{12}$ fps.

3) The exposure time is much longer than the framing time, and the spatial information is seriously confused.

The concept of degradation ratio[64] $D$, proposed in T-CUP article, is defined as the ratio of exposure time to framing time, which is the reciprocal of time information quality factor $g$ in the information theory of high-speed photography. Table 2 shows the four imaging rates and their corresponding framing time, exposure time, $D$ value, and g value; obviously, time information quality factor $g$ is far less than 1, and the information confusion is extremely serious. For example, its analysis value is not much with an aliasing ratio of 88.6% at the framing rate of 10×$10^{12}$ fps.

**Table 2 Relationship of main performance parameters**

4) Like the calculation of framing rate, the calculation of the effective frame number is also worth discussing. It is not appropriate to use the binned pixel size (2× or 1× CCD pixel size), but rather to use the imaging blur σ of the tube streak camera for framing. Due to the imaging blur σ of the current commonly used streak cameras is at least a dozen pixels, the effective frame number may vary by more than one order of magnitude. Moreover, the frame number energy principle should be considered [68].

## 5. Femtosecond imaging through coding on frequency spectrum plane (FRAME, FISI)

*5.1 Frequency recognition algorithm for multiple exposures (FRAME)*

FRAME is a fine femtosecond imaging technique to enable ultrafast 2D videography having spectroscopic compatibility with both high spatial and temporal resolution. Its coding is dependent on grating orientation-framing, which means that the framing formed by the difference of grating orientation in the spatial domain corresponding to different directions of its frequency spectrum on the frequency spectrum area of the system; and its decoding is carried out by superimposing structural codes onto CCD by a frequency recognition algorithm. The framing rate is determined by the difference in the delay time of each sub-light path [68]. In general, the upper limit of the rate is the reciprocal of the laser pulse width [32].

Figure 6 shows proof-of-principle demonstrations and schematics of the FRAME setup for visualizing the propagation of a femtosecond light pulse through a Kerr-sensitive medium. The fine feature is that it allows measuring at vastly different time scales, from second-to-microsecond to femtosecond. The main specifications are the framing rate of $5\times10^{12}$ fps corresponding to the framing time of 200 fs, the exposure time of 200 fs, and the spatial resolution of 15 lp/mm over a field-of-view of $7\times7$ mm$^2$, and the frame number of 4 imaged on the CCD of $1002\times1004$ pixels.

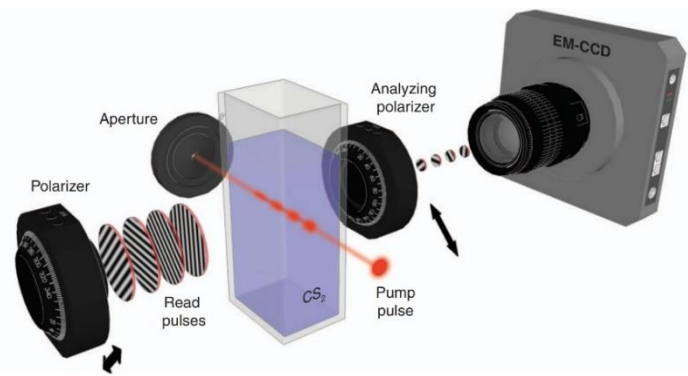

**Figure 6. FRAME setup.** Schematic diagram of FRAME setup for visualizing a femtosecond light pulse

As to frame number, notably, the energy principle should be considered because all image copies may need to share pixels and thus also share the camera's full well capacity (dynamic range). For example, if captured with a 16-bit sensor, the spatial structure of each of the eight image copies as a whole has a 13-bit resolution.

*5.2 optical frequency recognition imaging (FISI)*

In order to overcome the poor intrinsic spatial resolution and accuracy of the existing femtosecond imaging techniques owing to the limitations of streak cameras, compressed sensing algorithms, or other principle and structure reasons, Dr. Zhu Qifan [43],–presents the frequency domain integration sequential imaging (FISI), which encodes a transient dynamic by an inversed 4f system and decodes it using optical frequency recognition without the limitation of the frequency recognition algorithm. In an experiment on the process of an air plasma channel, FISI achieved shadow imaging of the channel with a framing rate of $1.26\times10^{13}$ fps and the maximum intrinsic spatial resolution of 108 lp/mm. With its outstanding intrinsic spatial resolution and high temporal resolution, FISI can probe both repeatable and unrepeatable ultrafast phenomena with high quality.

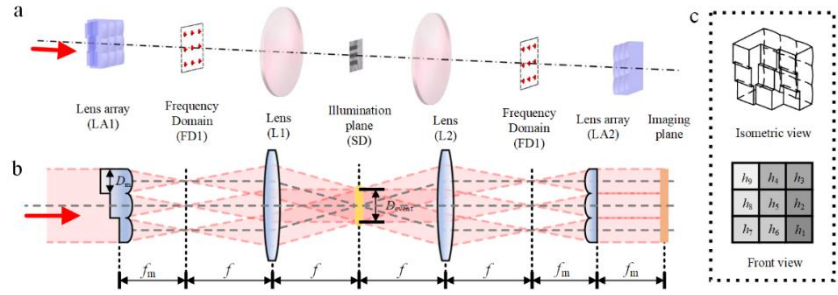

**Figure7. Schematic diagram of the optical system of FISI. A)** The 3D model of the FISI system including the frequency domains FD1 and FD2, the lenses L1 and L2 of the IF4 system, the spatial plane SD, the image plane IP, and the sub-lens in the lens array LA$_i$; **B)** light path diagram of FISI system; **C)** isometric and front view of the framing structure

FISI, consisting of an integrated framing structure, I4F system, and imaging structure, is a perfect system of femtosecond imaging because of the fine combination of an imaging system and illuminating system. The lens array unit integrates the framing function by modulating the illuminating laser beam and high-speed forming function by the lens array unit itself as a delay unit, the spatial resolution of the system is limited only by the lens and the size of the pixels in the detector, and the temporal resolution of it is determined only by the laser pulse width and the thicknesses and indexes of the glass sheet. The work of the important inversed 4f system (I4F) is shown in Figure7. 4f system, an optical information processing system consisting of two lenses, modulates spatial information in a frequency domain and can separate the spatial information into different frequency domain positions with small sizes to provide a possibility for integrating the framing structure. A point in a frequency domain FD1 is transferred into a spatial domain (SD) for illuminating an object placed in the center of SD, and the correspondent point carrying the object information imaged at the symmetry position in a frequency domain FD2, then imaged on the image plane IP. There are several symmetry points in FD2 with different times information. Finally, we can decode different times information in different positions.

The frequency recognition algorithm can solve a superimposed picture in the SD plane equivalent to the sampling plane, just as FRAME and holographic technologies. However, this algorithm's picture and sampling plane are limited by the frame and pixel sizes, which easily reduces the original spatial resolution. Additionally, as the detector's dynamic range is fixed, the number of frames and the dynamic range of each frame decrease, resulting in a decline in the resolution and accuracy. To address these two problems, the lens to perform Fourier transform and sampling in FD2 through the lens array LA can be used. As a result, the frames were exposed to different camera positions, the frame size or pixel size does not limit the optical frequency recognition, the spatial resolution is limited only by optical elements, and the camera's dynamic range does not affect the number of frames or accuracy.

The main specifications are the framing rate of $1.26 \times 10^{13}$ fps, the spatial resolution of 108 lp/mm, the exposure time of 35 fs, the frame number of 6, and the frame format of $4 \times 2.6$ mm$^2$. And the space bandwidth product of $1.2 \times 10^5$. So far, the femtosecond imaging system with the highest inherent spatial resolution and the higher space bandwidth product is a highlights recording, but increasing Schardin temporal-spatial information rate is always on the road.

## 6. Spectrally coded fs imaging (STAMP, SF-STAMP, IP-STAMP )

*6.1 Sequentially timed all-optical mapping photography (STAMP)*

The paper titled Sequentially timed all-optical mapping photography (STAMP), presented by Japanese scholar K. Nakagawa, was very attractive and exciting and referred to as the fastest speed photography owing to its maximum framing rate of $4.4 \times 10^{12}$ fps [41].

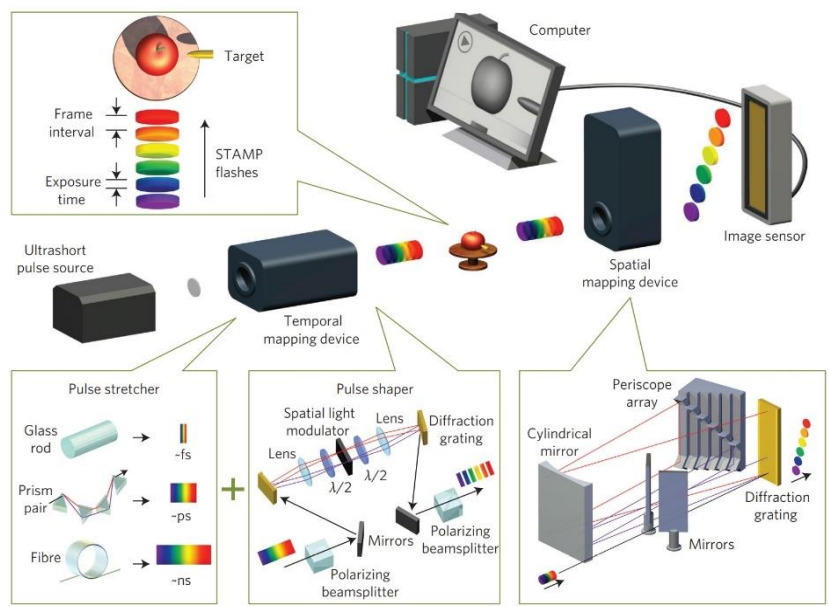

**Figure 8. Schematic diagram of STAMP**

As shown in Figure 8, STAMP consists of an ultrashort laser pulse, the temporal mapping device (TMD), a target, the spatial mapping device (SMD), the image sensor, and the computer. Its principle is transforming a time-varying spatial profile from the time domain to the spatial domain, using temporal dispersion and spatial separation for the transformation process. In STAMP, there are two key units: TMD splitting the ultrashort laser pulse into a series of discrete daughter pulses in different spectral bands, SMD separating the image-encoded daughter pulses and directing them towards different areas of the image sensor, optically and passively.

It is an advanced and ingenious system but too complicated in structure. Another problem is that its exposure time is far longer than framing time. Therefore, a severe overlap has occurred between two adjacent time frames. As for the exposure time of 733 fs, the framing time of 229 fs, and the framing rate of $4.4 \times 10^{12}$ fps, the spatial information aliasing ratio may be about 79%, and the temporal modification factor $g$ is very small, 0.37. Its analysis value is not much. The main specifications are the max.

The framing rate is $4.4×10^{12}$ fps, the data format is 450×450 pixels, the exposure time is 733 fs, and the frame number is 6.

An optimum design for STAMP in terms of temporal properties, including exposure time and framing rate, is presented [69]. Specifically, first derived master equations can be used to predict the temporal performance of a STAMP system and then analyze them to realize optimum conditions for serving as a general guideline for the camera parameters of a STAMP system with optimum temporal performance, that is, the ratio of the exposure time to the framing time at the minimum temporal resolution and the maximum possible total number of frames is found to be 1.4142. Nevertheless, this optimum result does not conform to the fundamental principle of high-speed imaging information theory.

*6.2 Sequentially timed all-optical mapping photography utilizing spectral filtering (SF-STAMP)*

For filling the insufficient of STAMP, SF-STAMP [70], as shown in Figure 9, is composed of a diffractive optical element (DOE) for generating array beams, a tilted band-pass filter (BPF) for spectrally resolving the diffracted array beams depending on their angles of the incident, and two Fourier transform lenses. They use a linearly frequency-chirped pulse and convert the wavelength to the time axis to realize single-shot ultrafast burst imaging. So that the wavelength-selected beams are optically inverse Fourier transformed by the second lens, and focused-2D multiple spectral bands images are simultaneously generated on the CCD plane. Its advantage is easy to realize more frame recording without a periscope array structure which is very complicated and hard manufactured.

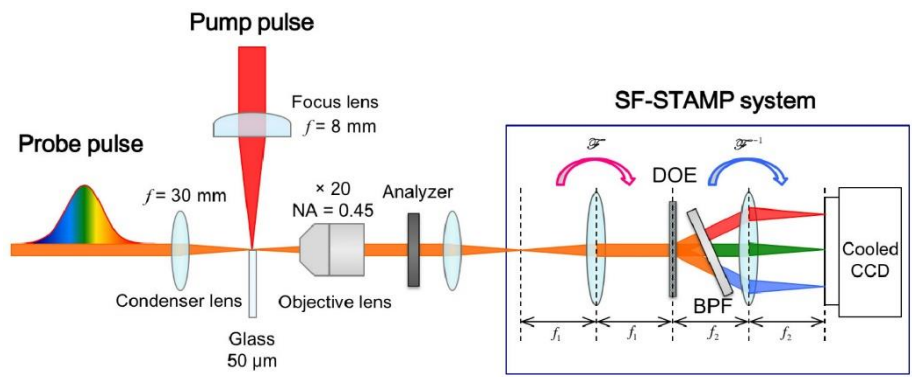

**Figure 9. SF-STAMP system.** Schematic diagram of microscopic SF-STAMP system for observation of ultrafast laser ablation dynamics.

The main specifications are the max. The framing rate of $2.09×10^{12}$ fps correspondent to the framing time of 477 fs, the exposure time of 477 fs, the data format of 740 × 480 pixels, and the frame number of 5; By used of a specially designed diffractive optical element (DOE) with a band-pass filter [71], the other method of SF-STAMP to increase the imaging frequency and frame number of STAMP system to $7.5×10^{12}$ fps with 25 frames. However, there is still a room for improvement in frame number.

To improve the optical throughput of SF-STAMP, lens array STAMP (LA-STAMP) [72] imaged the femtosecond laser-induced ablation process on a glass surface with $3.3 \times 10^{11}$ fps framing rate and ~4.4 μm spatial resolution in a 1.8 ns time window by introduced an integral field spectroscopy method using a micro-lens array to produce hyperspectral images. Although the frame number and the spatial resolution was limited to 7 and 4.4 μm in this prototype setup, it can be scaled to ~24 frames with a spatial resolution of ~1μm by designing an integral field spectroscopy with a fine pitch micro-lens array.

*6.3 Sequentially timed all-optical mapping photography utilizing spectral filtering on the imaging plane (IP-STAMP)*

A simple and compact design for sequentially timed all-optical mapping imaging for the femtosecond scale process, can be realized by four holograms corresponding to four different moments of the process and four chromatic channels of the special band-pass filter closed in front of CMOS. The imaging's frame interval and temporal resolution are determined by both the spectral separations and bandwidths of the channels in the filter and the group velocity dispersion induced by a dispersive element [73].

In this design, the different time delay with wavelength, based on material dispersion, is achieved by choosing N-BK7 optical glass as the dispersive element. As a result, the time intervals of 0, 319, 352, and 314 fs between the adjacent transmission peaks of 730, 770, 820, and 870 nm can be obtained, which means the average frame interval Δt ~330 fs. The key unit is a four-channel narrow band, the structure of which is that every sub-channel is arranged to be one set. Each sub-channel in every set has its own transmission wavelength marked with different colors, which different band-pass optical coatings can realize. The exposure time is limited by the duration of the sub-pulse, while the spatial resolution is dependent on the channel number and the number and sizes of the pixels of CMOS, and the relative delays determine the framing rate among the sub-pulses.

An up-dated IP-STAMP scheme with filter arrays on an image plane, named as CSMUP [74], has been reported as the following devices due to recent advances in microfilter technology. It can achieve single-shot real-time ultrafast imaging with a framing rate of about $2.50 \times 10^{12}$ fps. They imaged the dynamics of femtosecond laser ablation in silicon under a 400 nm femtosecond laser exposure with CSMUP. It is worth mentioning that its spatial resolving power is less than 833 nm, which provides a high spatial resolution imaging way to improve the efficiency and accuracy of femtosecond laser fabrication by a single-shot dynamic measurement of the interaction between the femtosecond laser and materials.

## 7. Raster-sampled femtosecond imaging (OPR)

All-optical high spatial-temporal resolution photography with raster principle at 2 trillion frames per second (OPR) is combining the sampling theory of raster and spectral-time coding technique through a grating, which has exhibited fine spatial-temporal resolution and framing rate/number, and more direct and faster reconstruction algorithm with high robustness, compared with compressed sensing algorithm. In a proof-of-principle demonstration, the time-resolved shadowgraph imaging of plasma filaments in the air and the transient scenes of plasma dynamics visualized in glass were recorded, respectively.

The transient scene can be expressed as discretized sequence depth frames $O(x,y,t_i)$, $I = 1,2,…,n$, and illuminated by the linearly chirped pulse $I(x,y,t(\lambda)_i)$ in which $t$ and $\lambda$ are mutually corresponding. The target is imaged on the plane of the microlens array to form a relay image, then a raster image, a finite set of sub-pupil image array and its intensity modulated by the target, can be formed by the microlens array.

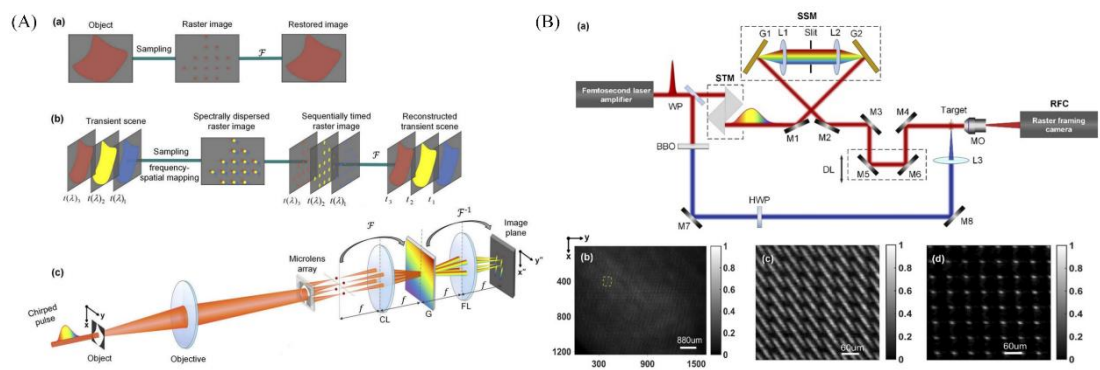

**Figure 10. OPR system. (A)** Schematic diagram of OPR. (a) Demonstrating sampling theory and Fourier reconstruction algorithm; (b) the operating principle; (c) The raster framing camera: collimating lens C, grating G, Fourier lens FL. **(B)** The experimental setup of OPR: (a) the ultrafast imaging in single-shot:, wedge plate WP, half wave plate HWP, , grating G1, G2, delay line DL, microscope objective MO; (b) The raw spectrally dispersed raster of probe pulse without an object; (c) The details in the yellow dotted box in Fig. 11(b); (d) The sub-bandwidth raster of a probe pulse

As shown in Figure 10, the OPR setup consists of a sequentially timed module (STM), spectral-shaping module (SSM), and raster framing camera (RFC), STM and SSM are used for linearly encoding frequency-time mapping and system calibration, respectively. The function of RFC is sampling target by microlens array and framing based on frequency-time-spatial positions conversion. The main specifications are the framing rate of $2\times10^{12}$ fps, the exposure time of 500 fs, the data format of 1626×1236 pixels, the frame number of 12, and the spatial resolution of 90 lp/mm on an objective plane. This imaging is novel in principle, available in the application, convenient in practice for femtosecond imaging, and has more room for improvement. For instance, using its flexible operating characteristics, the framing rate of $10\times10^{12}$ fps with more than 50 frames can be achieved by adjusting the designed parameters, such as diffraction grating with 300 lp/mm, 200 μm spacing of microlens arrays, shorter FWHM of a femtosecond pulse, and a larger detector array plane of the system.

### 8. OPA idler light femtosecond imaging (MOPA)

So far, it is the most reasonable imaging with multi-stages non-collinear optical parametric amplification idler light (MOPA), which can reach the highest effective framing rate with high spatial resolution in the atomic time scale imaging at a single shot. Using a laser-induced air plasma grating as a target, MOPA has realized 50 fs-resolved optical imaging with a spatial resolution of ~83 lp/mm and an effective framing rate of $15\times10^{12}$ fps, of course, which can be as high as $20\times10^{12}$ fps, depending on the exposure time. It has also successfully recorded an ultrafast optical lattice with its rotating speed up to $13.5\times10^{12}$ rad/s.

The experimental setup of MOPA, as shown in Figure 11, includes two parts: four stages of non-collinear optical parametric amplification idler imaging system and TUROL composed of Pulse stretcher PS, Spiral phase generator SPG, asymmetrical Michelson interferometer MI, Optical lenses L, Target TA, Quarter-wave plate QWP, Space-variant half wave plate SVHWP, mirrors M1~M4, Beam splitters BS1~BS2, and Time delay line TDL.

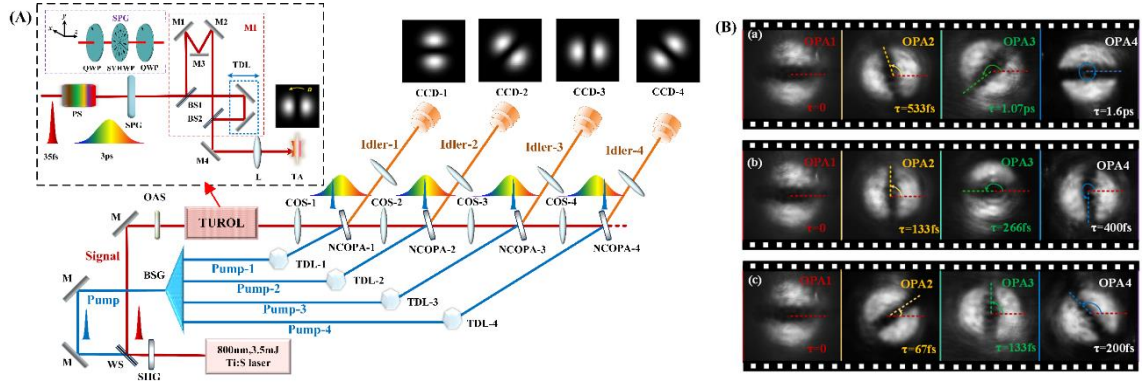

**Figure 11. MOPA system. (A)** The experimental setup of MOPA: M, Mirrors; WS, Wavelength separator; SHG, Second harmonic generator; NCOPA-1~NCOPA-4, Non-collinear optical parametric amplifiers; TDL-1~TDL-4, Time delay lines; BSG, Beam splitter group; COS-1~COS-4, Confocal optical systems; L1~L4, Optical lenses; CCD-1~CCD-4, Charge-coupled devices; TUROL unit. **(B)** Four frame images of the target ultrafast rotating optical lattice were recorded in the single-shot mode at the framing rate of $15\times10^{12}$ fps. (a),(b),(c), serial number of experiment.

Lots of experiments for increasing the intrinsic spatial resolution, the most important work for all of atomic time scale imaging, have been done, and the best result is the dynamic resolved power of 19 μm after eliminating spatial geometrical smearing influencing on spatial resolution with Type-II BBO crystal [75,76]. The experiments about single-shot MOPA femtosecond framing imaging include the framing imaging using a CW laser as a signal, using a chirped pulse as a signal for studying laser-induced plasma grating evolution at the framing rate of $10^{13}$ fps and using a chirped pulse as a signal for studying tunable ultrafast rotating optical ring-shape lattice (TUROL) at the framing rate of $15\times10^{12}$ fps [77,78,42].

This MOPA system can completely meet the second fundamental principle of atomic time scale imaging for realizing the optimal high-speed photography system because framing time depends on the relative time delay between pump beams, exposure time depending on pumping pulse width and of OPA, spatial resolution depending on pumping pulse (intensity, uniformity) and OPA crystal (thickness, category, and incision type ), and frame size depending on crystal size and pumping intensity, are irrelated with each other, and no limit of Heisenberg uncertainty principle. Moreover, the four key property specifications mentioned above can be further promoted. There is a bigger room to develop in it. We think that MOPA is a powerful tool for the potential applications in observing the acceleration process of laser wake field, spin orbit coupling process in hybrid halide perovskite materials, and phase transition process of semiconductor materials, manufacturing of two-dimensional materials and devices at the wafer level, the energy level jump ultrafast phenomenon of light fields, and so on.

## 9. Summarization and discussion

Since using a light-induced Kerr effect of $CS_2$ liquid as a shutter in about 1971, the great achievement in the atomic time scale imaging has been made, and in the early years of this decade, owing to the development of ultrafast laser technology, the time resolution of this imaging from picosecond into femtosecond region has been rapidly pushed. Notably, more new meaningful application scenario rather than the process of laser-induced plasma and Kerr effect may be realized in a broader hot research field. For instance, the dynamic process of light-field-driven currents in two-dimensional materials, the ultrafast Einstein-de Haas effect, ultrafast response of quantum states of light in a silicon chip and the process of topological transformation and phase transformation in two-dimensional materials, etc. Achievements described in this paper are innovative in principle, outstanding in performance, and typical in function and applying region; of course, the first proposed technique of this imaging should be emphasized.

In order to evaluate and develop atomic time scale imaging, it is necessary to establish and perfect the information theory on atomic time scale imaging based on the high-speed photography information theory established by Dr. H. Schardin. Our group has explored the initial works containing Schardin temporal-spatial information rate as the essential standard, the best optimum femtosecond imaging system free of Heisenberg uncertainty principle limit, and always pursuing the shorter exposure time and the fine analytical value spatial resolution and the bigger space-bandwidth product. It is worth mentioning that in recent research on CUP, the Schardin's equation has been referred to as space-bandwidth-time product (SBTP) [79], which is used to characterize the total amount of temporal-spatio information.

According to the information theory on atomic time scale imaging, the comparison among recent main imaging techniques of atomic time scale can be carried out as shown in Figure 12.

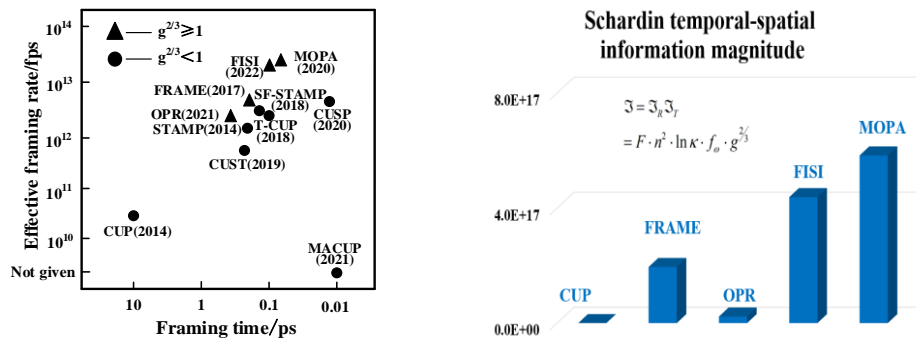

**Figure 12. Diagram of comparison among recent main imaging techniques of atomic time scale.** Effective imaging rate versus framing time(left), Schardin temporal-spatial information magnitude versus different technique (right)

In Figure 12 (right), some techniques, such as STAMP series, T-CUP series and so on, do not list their important parameters such as intrinsic spatial resolution and frame size in their papers, so they are not directly compared.

A question for discussion of the atomic time scale imaging to spring up may be explained as follows: deeply to explore new principles, available techniques, and reasonable structure, how to promote applied research of molecule/atom dynamics in photonic material, plasma physics, living cells, and neural activity, how to push the time scale from femtosecond to attosecond, and at last, how to catch the touching sight of electrons revolving around the nucleus, which is Gong Zutong Guess presented in 1980[80].

**Acknowledgments**
I am grateful to my group members of Dr. Hongyi Chen, Dr. Hu Long, and Prof. Xu Shixiang for their excellent work in atomic time scale imaging which makes my writing this paper to be possible. This paper belongs to Perspectives and Review articles.

**Funding:** This work was supported by financial supports from National major scientific research instrument research projects National Natural Science Foundation of China (61827815).

**Conflicts of interests:** The author declares that there is no conflict of interest regarding the publication of this article.

**Data Availability:** Data underlying the results presented in this paper are not publicly available at this time but may be obtained from the authors upon reasonable request.